\documentclass{ws-procs10x7}

\usepackage{epsfig}

\def\Journal#1#2#3#4{{\it #1} {\bf #2}, #3 (#4)}

\def\PLB{{\em Phys. Lett.}   {\bf B}}

\def\PRD{{\em Phys. Rev.}    {\bf D}}

\def\EJC{{\em Eur. Phys. J.} {\bf C}}

\newcommand{\pom}{{I\!\!P}}
\newcommand{\xpom}{x_\pom}

\begin{document}

\title{DIFFRACTIVE DIS CROSS SECTIONS AND PARTON DISTRIBUTIONS}

\author{F.-P. SCHILLING (H1 Collaboration)}

\address{
Physics Department, CERN, CH-1211 Geneva 23, Switzerland\\
E-mail: frank-peter.schilling@cern.ch
}

\twocolumn[\maketitle\abstract{
Highlights are presented mainly from two recent measurements of the
diffractive Deep Inelastic Scattering cross section at HERA. In the
first, the process $ep\rightarrow eXp$ is studied by tagging the
leading final state proton. In the second, events of this type are
selected by requiring a large rapidity gap devoid of hadronic activity
in the proton direction. The two measurements are compared in detail
and the kinematic dependences are interpreted within the framework of
a factorisable diffractive exchange. Diffractive parton distributions
are determined from a next-to-leading order QCD analysis of the large
rapidity gap data, which can be applied to the prediction of
diffractive processes, also at the TEVATRON and the LHC.  
}]


\section{Introduction}

This report\footnote{Talk presented at ICHEP 2006, Moscow}
summarises recent results on measurements of the diffractive
deep-inelastic scattering (DIS) cross section obtained with the H1
detector at the HERA $ep$ collider, in particular from two recent
publications\cite{f2d4,f2d3}.  The measurements cover an unprecedented
kinematic range of photon virtualities $3.5 < Q^2 < 1600
\rm\ GeV^2$ with unprecedented precision (5\% statistical, 5\%
systematic and 6\% normalisation errors in the best-measured region).

In the first paper\cite{f2d4}, the Forward Proton Spectrometer (FPS)
is used to detect and measure the four-momentum of the outgoing proton
in the process $ep\rightarrow eXp$. This selection method has the
advantages that the proton unambiguously scatters elastically and that
the squared four-momentum transfer at the proton vertex $t$ can be
reconstructed. However, the available statistics are limited by the
FPS acceptance.  A high statistics sample of diffractive DIS events is
selected on the basis of a large rapidity gap (LRG) in the outgoing
proton direction, as described in the second paper\cite{f2d3}. The
measured process is $ep\rightarrow eXY$ where $Y$ corresponds to any
baryonic state with mass $M_Y<1.6 \rm\ GeV$.  Together, the FPS and
LRG data provide a means of studying inclusive diffraction as a
function of all relevant variables. In addition to $t$ and the usual
DIS variables $x$ and $Q^2$, measurements are made as a function of
the fractional proton longitudinal momentum loss $\xpom$ and of
$\beta=x/\xpom$, which corresponds to the fraction of the exchanged
longitudinal momentum carried by the quark coupling to the virtual
photon.

The data exhibit a remarkable consistency with proton vertex
factorisation\cite{schlein}, where the dependences on $\xpom$, $t$ and
$M_Y$ describing the proton vertex are completely independent of
$\beta$ and $Q^2$, which describe the hard interaction with the
photon. The dependences on $\xpom$ and $t$ can then be expressed in
terms of an {\em effective pomeron flux} of colourless exchange,
whilst the $\beta$ and $Q^2$ dependences can be interpreted in terms
of {\em diffractive parton distributions} (DPDFs), which describe the
partonic structure of that exchange\cite{collins}.

\begin{figure*}[t]
\centering
\epsfig{file=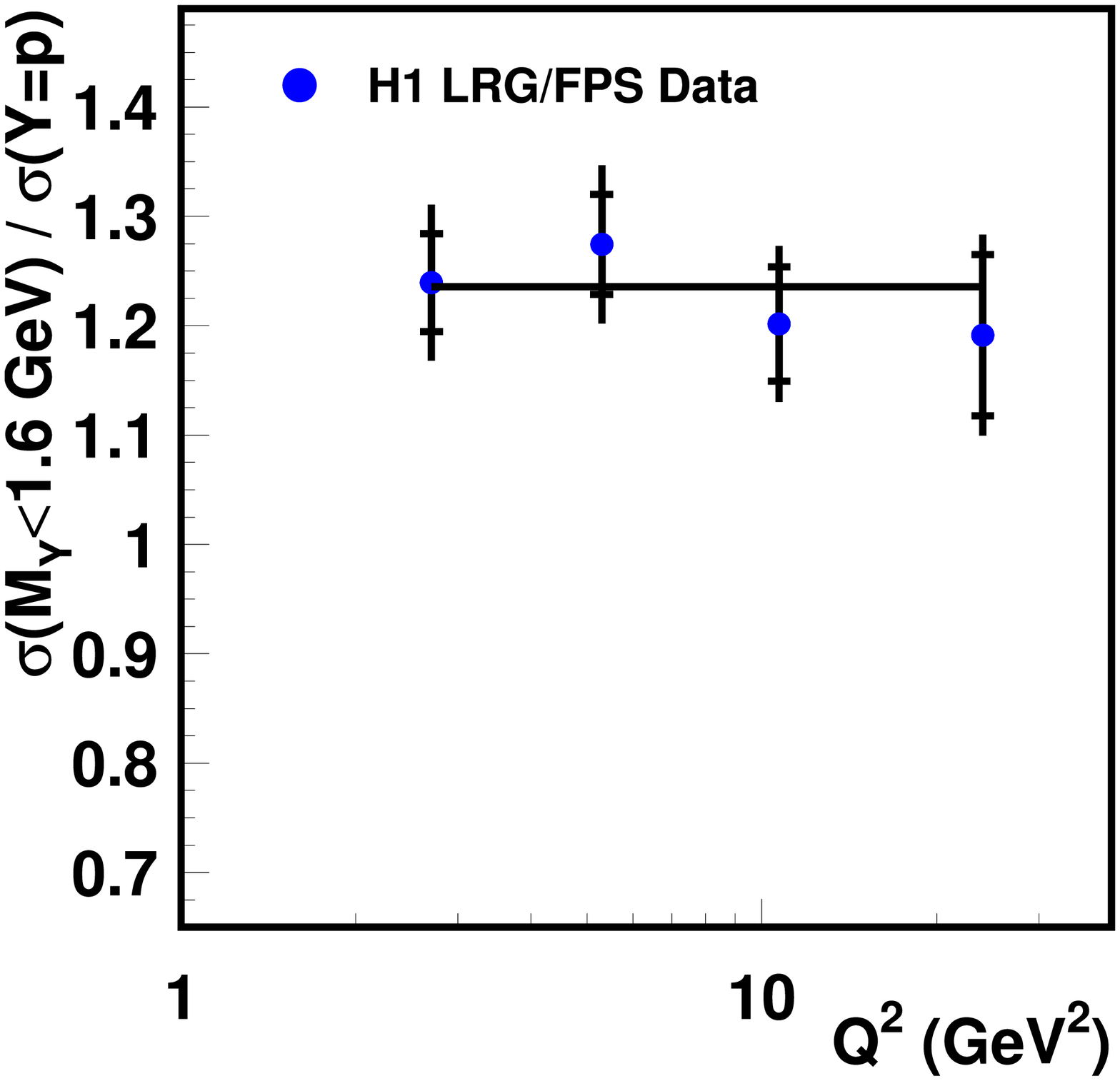,width=0.3\linewidth}
\epsfig{file=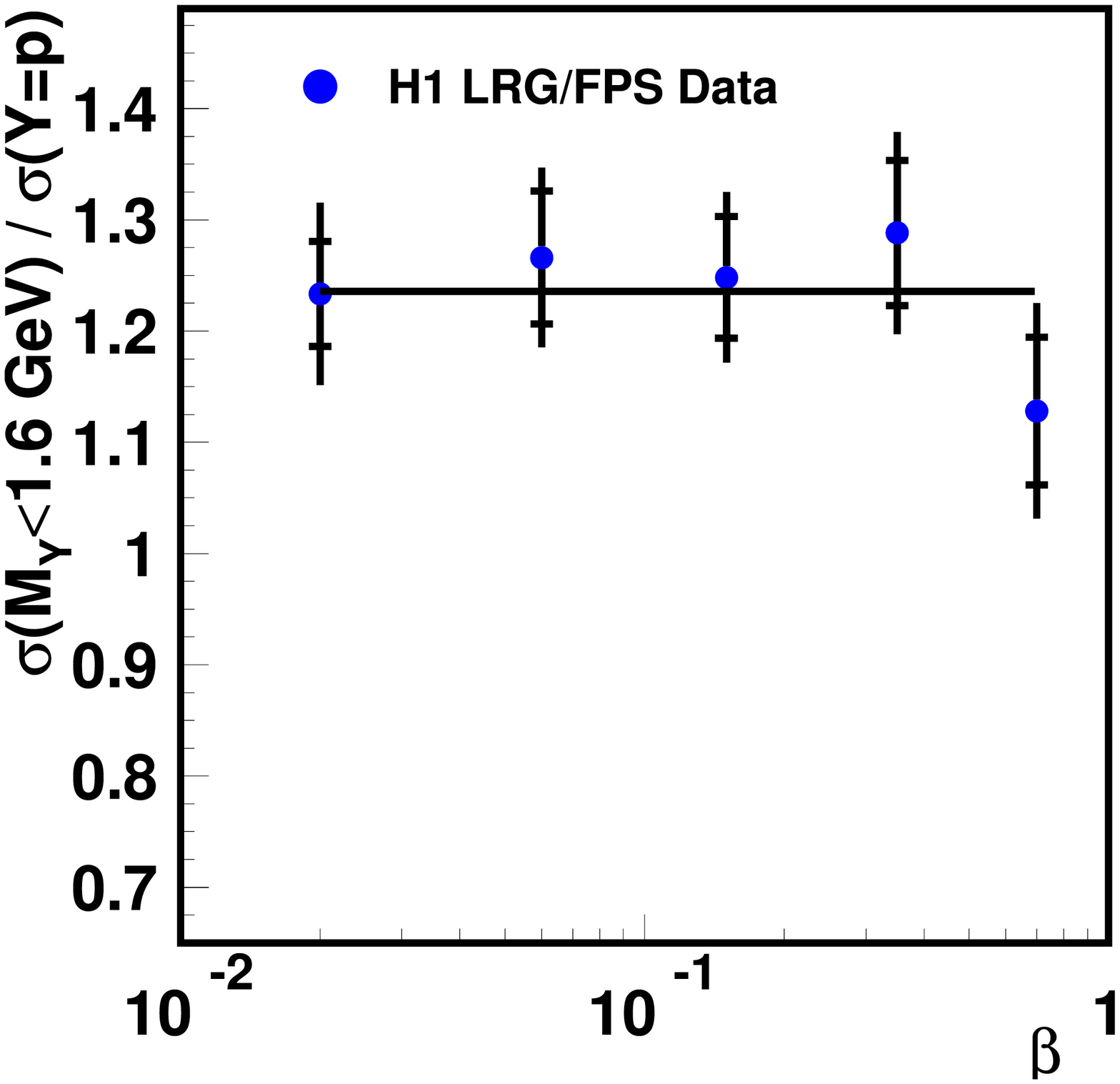,width=0.3\linewidth}
\epsfig{file=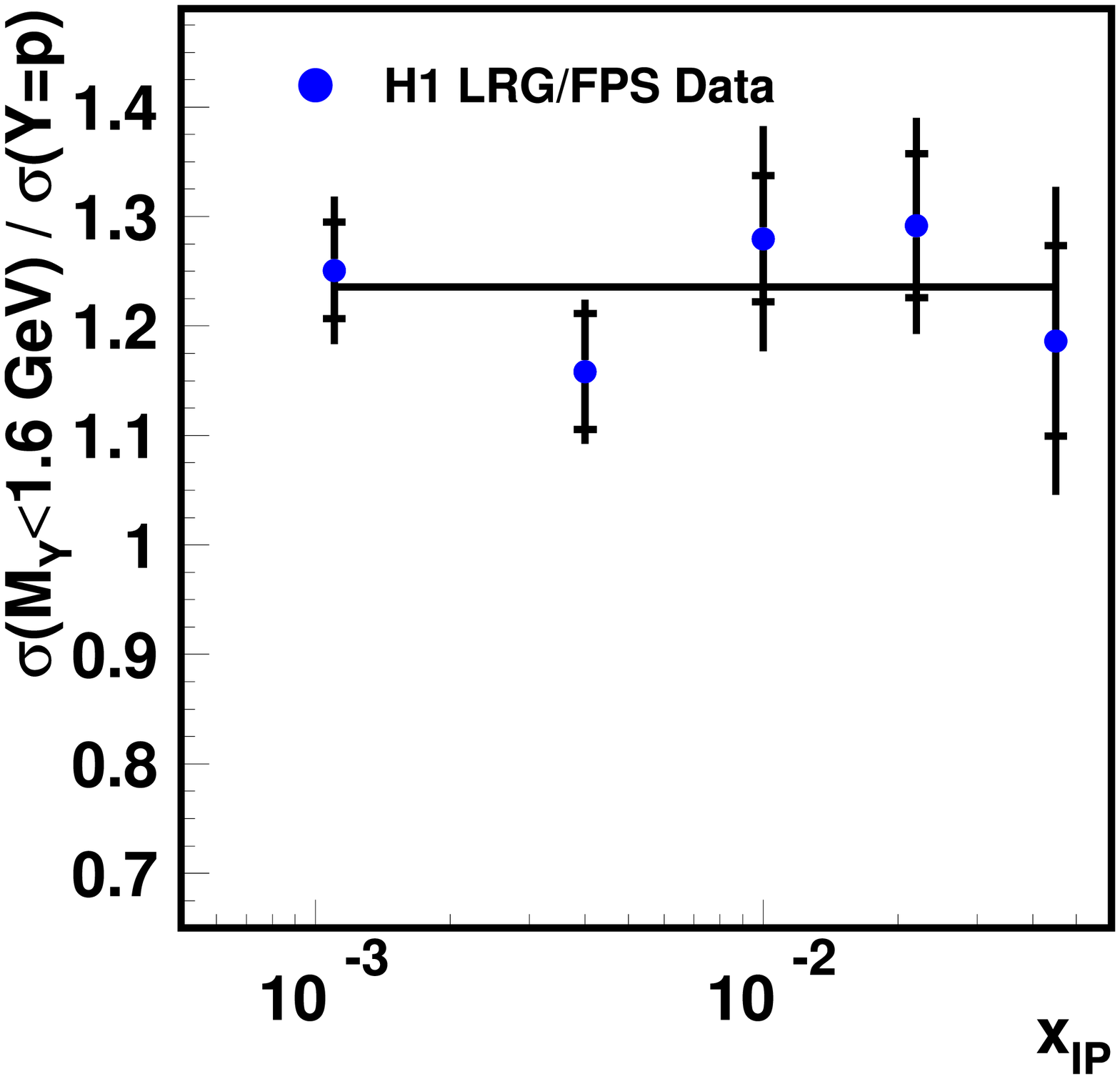,width=0.3\linewidth}
\caption{
The ratio of the cross section for $M_Y < 1.6 \rm\ GeV$ and $|t| < 1
\rm\ GeV^2$ (LRG data) to that for $Y = p$ and $|t| < 1 \rm\ GeV^2$
(FPS data) as a function of $Q^2$, $\beta$ and $\xpom$, averaged over
the other variables. A $13\%$ normalisation uncertainty is not shown.
}
\label{fig:ratio}
\end{figure*}

Only a short summary of a few highlights is possible here. Much more
detail, including the multi-differential cross section measurements
themselves, can be found in\cite{f2d4,f2d3}. The first charged current
diffractive measurement, as well as ratios of the diffractive and
inclusive cross sections, are also presented in\cite{f2d3}, but not
covered here. New H1 diffractive DIS measurements with increased
statistical precision, but in a limited kinematic range, were also
presented\cite{h1newdata}.


\section{Comparison between Data Sets}

Since the LRG and FPS data sets are statistically independent and have
very different systematics, the two measurements constitute a powerful
mutual cross-check. Compatibility between them is established in
detail by performing $t$-integrated measurements by both techniques
with identical binning and forming the ratio of the two measurements
for each ($Q^2$, $\beta$, $\xpom$) point\cite{f2d4}.

The dependences of this ratio on each kinematic variable individually
is shown in Fig.~\ref{fig:ratio} after taking statistically weighted
averages over the other two variables. Within the uncertainties of
typically 10\% per data point, there is no significant dependence on
$\beta$, $Q^2$ or $\xpom$. The ratio of overall normalisations, LRG /
FPS, is $\sigma(M_Y < 1.6 {\rm\ GeV}) / \sigma(Y=p) = 1.23 \pm0.03
{\rm\ (stat.)} \pm 0.16 {\rm\ (syst.)}$, consistent with predictions
for the proton-elastic cross section and the proton dissociation cross
section with $M_Y < 1.6 \rm\ GeV$\cite{f2d4}. The FPS data are also
consistent with the corresponding measurement obtained with the ZEUS
Leading Proton Spectrometer\cite{lps}.

\begin{figure}[t]
\centering
\epsfig{file=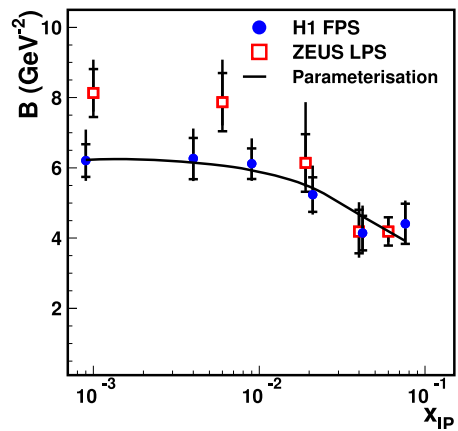,width=0.9\linewidth}
\caption{
Measurements of the slope parameter $B$ by H1 and ZEUS and a
parameterisation of the H1 data as used to describe the pomeron flux
factor.  }
\label{fig:tslope}
\end{figure}


\section{Dependences on $\xpom$ and $t$ \label{sec:xpomtdep}}

The $t$ dependences of diffractive cross sections are commonly
parameterised with an exponential, $d\sigma / dt \sim e^{Bt}$. The
values of $B$ resulting from such fits to the FPS data are shown as a
function of $\xpom$ in Fig.~\ref{fig:tslope}. At low $\xpom$, the data
are compatible with a constant slope parameter, $B\simeq 6 \rm\
GeV^{-2}$.  In a Regge approach with a single linear exchanged pomeron
trajectory, $\alpha_\pom(t) = \alpha_\pom(0) + \alpha'_\pom t$, the
slope parameter decreases with increasing $\xpom$ according to $B =
B_0 - 2\alpha'_\pom \ln \xpom$. The low $\xpom$ data thus favour a
small value of $\alpha'_\pom \simeq 0.06 \rm\ GeV^{-2}$, though
$\alpha'_\pom \simeq 0.25$, as obtained from soft hadronic
interactions, cannot be excluded.

The $\xpom$ dependences of both measurements are interpreted in terms
of effective pomeron intercepts. The two results are consistent, the
more precise value of $\alpha_\pom(0) = 1.118 \pm 0.008 {\rm\ (exp.)} 
^{+0.029}_{-0.010} {\rm\ (model)}$ coming from the LRG data. The
dominant error arises from the strong positive correlation between
$\alpha_\pom(0)$ and $\alpha'_\pom$, such that $\alpha_\pom(0)$
increases to around $1.15$ if $\alpha'_\pom$ is set to $0.25 \rm\
GeV^{-2}$ rather than $0.06 \rm\ GeV^{-2}$.  The extracted
$\alpha_\pom(0)$ is slightly higher than the `soft pomeron' value of
$\alpha_\pom(0)\simeq1.08$, obtained from long distance hadronic
interactions.  The values of both $\alpha_\pom(0)$ and $\alpha'_\pom$
describing diffractive DIS are compatible with the results obtained
for soft exclusive photoproduction of $\rho^0$
mesons\cite{krueger}. This similarity supports the picture of
diffractive DIS as probing the structure of a `soft' pomeron. `Hard'
perturbative two gluon exchange contributions are likely to be small,
as is also suggested by the lack of a signal for exclusive dijet
production\cite{h1jets}.

\begin{figure}[t]
\centering
\epsfig{file=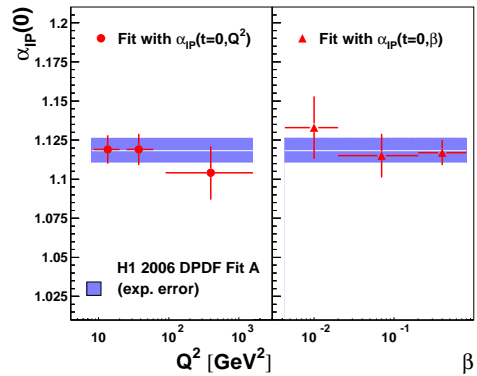,width=0.9\linewidth}
\caption{
Effective pomeron intercept $\alpha_\pom(0)$ as extracted from the LRG
data, showing no significant variation with $Q^2$ or $\beta$.  }
\label{fig:a0pom}
\end{figure}

Further analysis in which either the slope $B$ or the intercept
$\alpha_\pom(0)$ is allowed to vary with $\beta$ or $Q^2$ shows no
significant dependences (Fig.~\ref{fig:a0pom}), confirming the
validity of proton vertex factorisation for the present data. This
contrasts with the $Q^2$ dependent effective pomeron intercept
extracted in a Regge approach to inclusive low $x$ proton structure
function data, as studied in detail via the ratio of diffractive to
inclusive cross sections in\cite{f2d3}.

\section{Dependences on $\beta$ and $Q^2$: Diffractive Parton Densities}

In\cite{f2d3}, the cross section is presented differentially in
$\beta$, $Q^2$ and $\xpom$. After dividing out the $\xpom$ dependence
using a flux factor with parameters obtained as described in section
\ref{sec:xpomtdep}, the results from different $\xpom$ values are
compatible, as expected where proton vertex factorisation holds. 

\begin{figure}[t]
\centering
\epsfig{file=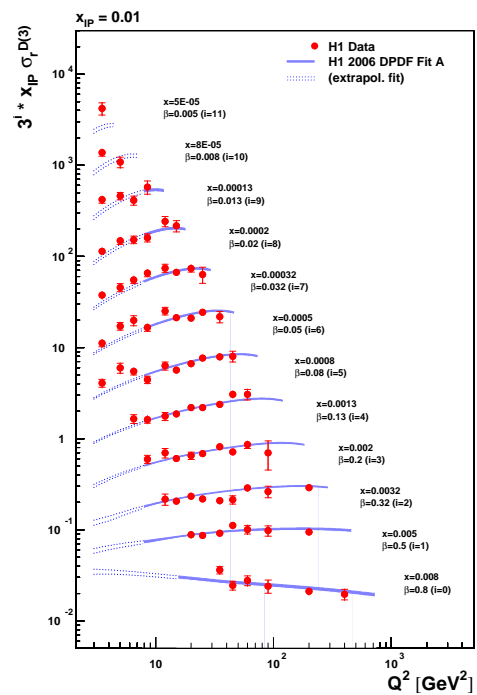,width=0.9\linewidth}
\caption{
$Q^2$ dependence of the diffractive DIS cross section at $\xpom=0.01$
for different values of $\beta$.  }
\label{fig:dataq2}
\end{figure}

\begin{figure}[t]
\centering
\epsfig{file=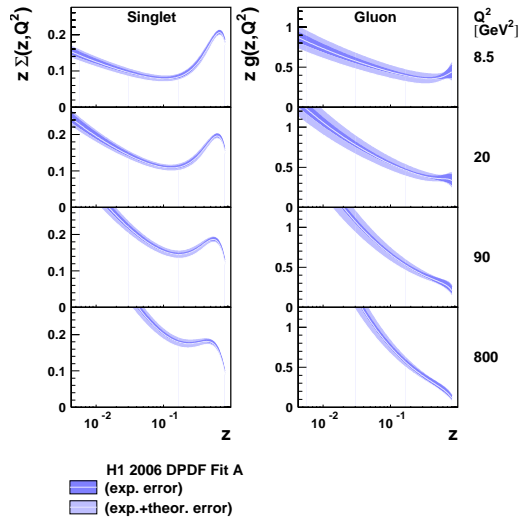,width=0.99\linewidth}
\caption{
Quark singlet and gluon distributions from the NLO QCD fit `H1
2006 DPDF Fit A', as a function of the momentum fraction $z$ carried
by the relevant parton.
}
\label{fig:pdfs}
\end{figure}

The $\beta$ and $Q^2$ dependences (Fig.~\ref{fig:dataq2}) of the data
are interpreted in a next-to leading order (NLO) DGLAP QCD
fit\cite{f2d3} in order to extract DPDFs. For the first time,
experimental and theoretical uncertainties are evaluated for these
partons. The results are shown in Fig.~\ref{fig:pdfs}.

\begin{figure}[t]
\centering
\epsfig{file=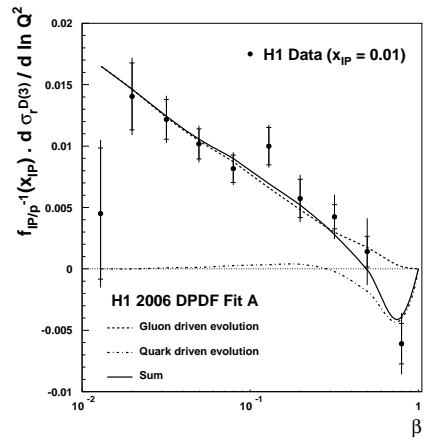,width=0.9\linewidth}
\caption{
Logarithmic $Q^2$ 
derivatives of the diffractive cross section at fixed $\xpom=0.01$
as a function of $\beta$. 
}
\label{fig:logq2deriv}
\end{figure}

The singlet quark density is very closely related to the measured
diffractive cross section and is thus well constrained, with a typical
error of $5\%$. According to the DGLAP evolution equations, the
logarithmic $Q^2$ derivative (shown in Fig.~\ref{fig:logq2deriv} for
$\xpom=0.01$) contains contributions due to the splittings
$g\rightarrow q\bar{q}$ and $q\rightarrow qg$, convoluted with the
diffractive gluon and quark densities, respectively. The derivative is
determined almost entirely by the diffractive gluon density up to
$\beta\simeq 0.3$. The large positive $\ln Q^2$ derivatives in this
region can thus be attributed to a large gluonic component in the
DPDFs. For $\beta>0.3$, the contribution to the $Q^2$ evolution from
quark splittings $q\rightarrow qg$ becomes increasingly important and
the derivatives become less sensitive to the gluon density. The gluon
density is thus known to around $15\%$ at low $\beta$, with an
uncertainty that grows quickly for $\beta>0.3$.

These DPDFs provide important input to final state measurements such
as those involving jets and charm quarks\cite{kapishin}, which may
also provide important additional constraints\cite{h1jetsprel} on the
gluon at high $\beta$.  Integrated over $\beta$, the gluon density
carries around $70\%$ of the total momentum. A similar fraction of the
total proton momentum is carried by the inclusive gluon density in the
low $x$ region where valence quark effects are small. This similarity
of the ratio of quarks to gluons in the DPDFs and the inclusive proton
parton densities is reflected\cite{f2d3} in a ratio of the two cross
sections which, to good approximation, is flat as a function of $Q^2$
at fixed $x$ and $\xpom$.

The DPDFs may also be used in calculations of diffractive cross
sections at the TEVATRON as well as the LHC.


\end{document}